\documentstyle[12pt]{article}
\def\sq{\hbox {\rlap{$\sqcap$}$\sqcup$}}
\def\sq{\hbox {\rlap{$\sqcap$}$\sqcup$}}
\def\R{ {\rm R \kern -.31cm I \kern .15cm}}
\def\C{ {\rm C \kern -.15cm \vrule width.5pt \kern .12cm}}
\def\Z{ {\rm Z \kern -.27cm \angle \kern .02cm}}
\def\N{ {\rm N \kern -.26cm \vrule width.4pt \kern .10cm}}
\def\1{{\rm 1\mskip-4.5mu l} }
\def\lsim{\raise0.3ex\hbox{$<$\kern-0.75em\raise-1.1ex\hbox{$\sim$}}}
\def\gsim{\raise0.3ex\hbox{$>$\kern-0.75em\raise-1.1ex\hbox{$\sim$}}}
\def\noi{\noindent}

\def\beq{\begin{equation}}   \def\eeq{\end{equation}}
\def\bea{\begin{eqnarray}}  \def\eea{\end{eqnarray}}
\def\nn{\nonumber}
\def\noi{\noindent}
\def\beeq{\begin{eqnarray}} \def\eeeq{\end{eqnarray}}
\newcommand\mysection{\setcounter{equation}{0}\section}
\renewcommand{\theequation}{\thesection.\arabic{equation}}
\newcounter{hran} \renewcommand{\thehran}{\thesection.\arabic{hran}}

\def\bmini{\setcounter{hran}{\value{equation}}
   \refstepcounter{hran}\setcounter{equation}{0}
   \renewcommand{\theequation}{\thehran\alph{equation}}\begin{eqnarray}}

\def\bminiG#1{\setcounter{hran}{\value{equation}}
\refstepcounter{hran}\setcounter{equation}{-1}
\renewcommand{\theequation}{\thehran\alph{equation}}
\refstepcounter{equation}\label{#1}\begin{eqnarray}}

\def\emini{\end{eqnarray}\relax\setcounter{equation}{\value{hran}}\renewcommand{\theequation}{\thesection.\arabic{equation}}}

\begin{document}
\centerline{\Large\bf Blown-up  p-Branes and the Cosmological Constant}
\vskip 1 truecm

\begin{center}
{\bf Ulrich Ellwanger}\footnote{E-mail :
Ulrich.Ellwanger@th.u-psud.fr}\par \vskip 5 truemm

Laboratoire de Physique Th\'eorique\footnote{Unit\'e
Mixte de Recherche - CNRS - UMR 8627}\par
  Universit\'e de Paris XI, B\^atiment
210, F-91405 Orsay Cedex, France\par \vskip 5 truemm
\end{center}
\vskip 2 truecm

\begin{abstract}
We consider a blown-up 3-brane, with the resulting geometry $R^{(3,1)}
\times S^{(N-1)}$, in an infinite-volume bulk with $N > 2$ extra
dimensions. The action on the brane includes both an Einstein term and
a cosmological constant. Similar setups have been proposed both to
reproduce 4-d gravity on the brane, and to solve the cosmological
constant problem. Here we obtain a singularity-free solution to
Einstein's equations everywhere in the bulk and on the brane, which
allows us to address these question explicitely. One finds, however,
that the proper volume of $S^{(N-1)}$ and the cosmological constant on
the brane have to be fine-tuned relatively to each other, thus the
cosmological constant problem is not solved. Moreover the scalar
propagator on the brane behaves 4-dimensionally over a
phenomenologically acceptable range only if the warp factor on the
brane is huge, which aggravates the Weak Scale -- Planck Scale
hierarchy problem.
  \end{abstract}

\vskip 3 truecm
\noi LPT Orsay 03-11 \par
\noi March 2003 \par

\newpage
\pagestyle{plain}
\baselineskip 20pt
\mysection{Introduction}
\hspace*{\parindent}
Recently it has been claimed that the cosmological constant problem can
be solved in brane worlds with $N > 2$ infinitely large extra
dimensions with an extra Einstein term localized on the brane
\cite{1r,2r}. Then the graviton propagator is soft (massive) beyond
some scale $r \ \gsim\ r_c$, and consequently the gravitational field
(the FRW scale factor $a(t)$) does not necessarily ``react'' to sources
that are smooth at scales $\gsim\ r_c$ such as a cosmological constant.
The success of the cosmological standard model, on the other hand,
requires $r_c \ \gsim\ H_0^{-1}$ where $H_0$ is the Hubble constant
today. \par

The setup in \cite{1r,2r} assumes a vanishing cosmological constant in
the bulk, which can possibly be motivated by some unbroken
supersymmetry (and $R$-parity) in the bulk. A 4-dimensional behaviour
of gravity, hopefully over a large range of scales, is then due to an
additional Einstein term on the brane [3-5]. The (classical)
brane-to-brane graviton propagator actually suffers from a short
distance singularity for $N>2$, which requires an UV regularization in
the form of ``blowing up'' the transverse size of the 3-brane, higher
derivative terms in the bulk or a momentum cutoff [6-12]. \par

A solution of the cosmological constant problem is achieved only if one
includes, in addition, an arbitrary (not fine-tuned) cosmological
constant on the brane. This induces a non-trivial gravitational field
in the bulk surrounding the brane, which generically exhibits a naked
singularity \cite{13r}. Again a regularization of this singularity
requires a blowing-up of the brane \cite{14r,10r} (or higher
derivative terms \cite{12r}). \par

In \cite{1r,2r} it has been argued that allowing the brane to inflate
-- with a phenomenologically acceptable small acceleration rate --
removes the naked singularity. However, no explicit solution of the
Einstein equations was given, and the proposed scenario supposedly
requires at least fine-tuned initial conditions.\par

In the present paper we present a (singularity free) solution of
Einstein's equation, which corresponds to a blown-up 3-brane in $N > 2$
extra dimensions. In contrast to smoothed-out branes considered
elsewhere Einstein's equations are satisfied everywhere: Not only in
the bulk for $r > r_b$ (where $r_b$ is the radius of the blown-up
brane) but also for $r < r_b$. This is made possible through a
particular choice for the profile function $f(r)$, which describes the
distribution of the brane tension as a function of $r$ (which was
concentrated at $r = 0$ originally): We assume that the complete brane
tension (i.e. the brane cosmological constant) is concentrated at the
surface $r = r_b$ of the blown-up brane, i.e. $f(r) \sim \delta (r -
r_b)$. The shape of the brane in the $N$ extra dimensions is then the
one of a sphere $S^{(N-1)}$. Although the distribution of the brane
tension is still $\delta$-function-like in the radial direction, this
does not imply a singular gravitational field in the bulk (or on the
brane) since one has just an effective co-dimension one problem. \par

A similar setup has already been considered in \cite{8r,10r}. In
\cite{10r}, however, the metric in the bulk ``inside'' the sphere (for
$r < r_b$) was assumed to be constant, and no consistent solution to
the junction conditions could be found.\par

Here we take instead, for $r < r_b$, a copy of the known metric
``outside'' \cite{13r,14r} after an inversion $r \to r_b^2/r$. This
configuration has a $\Z_2$ symmetry: Changing the radial coordinate from
$r$ to $y$ by $r = r_b \exp (y/r_b)$ our setup is invariant under a
reflection at $y = 0$; the surface of $S^{(N-1)}$ is situated at $y=0$.
Now the notions ``inside'' and ``outside'' have actually lost their
meaning, since the space ``inside'' is just a copy of the space
``outside'' and shares with it its infinite volume. \par

Nevertheless this corresponds to a regularization (``blowing-up'') of
an infinitely thin 3-brane: Fields living on the brane with its world
volume $R^{(3,1)} \times S^{(N-1)}$ (where $R^{(3,1)}$ is our
4-dimensional Minkowski space) just see the finite world volume of
${\cal O}(r_b^{N-1})$ of $S^{(N-1)}$ independently from the infinite
volumes ``inside'' and ``outside'', and their Kaluza-Klein modes on
$S^{(N-1)}$ become infinitely heavy for $r_b \to 0$. \par

Since we have explicit singularity-free solutions of Einstein's
equations and junction conditions, we can ask whether the brane can
support an arbitrary tension (cosmological constant) without inflation,
i.e. without a naked singularity in the static case. It turns out,
however, that the brane tension and the volume of the blown-up brane
have to be relatively fine-tuned. This follows from the mere number of
junction conditions to be satisfied, hence we have all reasons to
believe that this result is generic and independent from the particular
setup considered here (a profile function concentrated on the
boundary). The Einstein term on the brane does not play a particular
role for this result, it just modifies the required relation between
the brane tension and its volume on $S^{(N-1)}$. \par

Next we can ask whether the gravitational propagator on the brane
behaves 4-dimensionally over a phenomenologically acceptable range of
scales or momenta $p^2$. To this end we study the scalar propagator on
the brane in the present gravitational background, with coefficients of
the kinetic terms in the bulk and on the brane identical to the ones of
the Einstein terms. This does not yet allow us to study the tensor
structure of the graviton propagator, but a $p^{-2}$
behaviour of the scalar propagator over a large range of $p^2$ is
already a necessary condition for an acceptable behaviour of gravity.
\par

The result is that, one the one hand, a $p^{-2}$ behaviour of the 
scalar propagator over a phenomenologically acceptable range is
possible, for a certain range of values of the gravitational constant
$M_D$ in the $D$-dimensional bulk and the volume of $S^{(N-1)}$.
However, for the corresponding values of these parameters the warp
factor $A$ on the brane is huge:
\beq
\label{1.1e}
A > 10^{15 \sqrt{{N-1 \over N + 2}}} \ .
\eeq

This behaviour of the warp factor is opposite to the one proposed in
\cite{15r} as a solution to the Weak Scale -- Planck Scale hierarchy 
problem. A scenario with many {\it infinite} extra dimensions, and 4-d
gravity due to an (induced) Einstein term on the brane, is thus a
priori difficult to realize.\par

On the other hand recall that, for a 4-dimensional behaviour of gravity
over a phenomenologically acceptable range, the setup in [2] requires
(for $N > 1$)

\beq
\label{1.2e}
\left ( {M_{Pl} \over M_D} \right )  > 10^{30} \ .
\eeq

Here we find that the required hierarchy between the 
gravitational constants $M_D$ and $M_{Pl}$ is less dramatic, if one
allows the proper volume of $S^{(N-1)}$ to be as small as
$M_{Pl}^{1-N}$ (instead of $M_{D}^{1-N}$): Then a $p^{-2}$ behaviour of
the propagator over a phenomenologically acceptable range
requires only

\beq
\label{1.3e}
\left ( {M_{Pl} \over M_D} \right )^{N+2}  > 10^{30} 
\eeq

\noi which, for $N$ large, is easier to realize than eq. (\ref{1.2e}).
\par

In the next section we present our setup more explicitly, and solve the
combined Einstein equations in the bulk and junction conditions on the
blown-up brane. The existence of static singularity-free
(non-supersymmetric) $(D-2)$-brane configurations in $D$-dimensions,
with $N - 1$ dimensions of the brane compactified on $S^{(N-1)}$, is
quite remarkable in view of the naked singularities encountered
generally \cite{13r} and the negative result in \cite{10r}. As noted
above, however, a relative fine-tuning of the input parameters is
required, thus the cosmological constant problem is not solved. \par

In Section III we study the scalar propagator in this background.
Although the full problem (the corresponding wave equation for all
non-zero modes) cannot be solved explicitely \cite{14r}, the partial
results in \cite{14r} can be used here as well and allow to obtain the
essential features of the $p^2$-dependence of the propagator. Then we
derive the phenomenological constraints on the input parameters cited
above and conclude.

\mysection{Singularity Free Blown-up p-Branes}
\hspace*{\parindent}
The original scenario consists in an infinitely thin 3-brane in a $D =
4 + N$ dimensional bulk, i.e. with $N$ extra dimensions. Our
conventions for coordinates and indices are: $x^{\mu}$, $\mu = 0 \dots
3$, are the coordinates of flat Minkowski space $R^{(3,1)}$ along the
3-brane, with the metric $\eta_{\mu\nu} = \ {\rm diag} (-1, 1, 1, 1)$.
$y^i$, $i = 1 \dots N$, denote the coordinates of the extra dimensions.
Indices $A, B = 0 \dots D-1$ run over all $D = 4 + N$ coordinates. \par

The original action includes an Einstein term in the bulk, and an
Einstein term plus a cosmological constant on the brane situated at
$y^i = 0$:

\beq
\label{2.1e}
S = \int d^4x\ d^Ny \left \{ \sqrt{-g^{(D)}} {1 \over 2 \kappa_D}
R^{(D)} + \sqrt{- g^{(4)}} \delta^N(y) \left [ {1 \over 2 \kappa_4}
R^{(4)} + \Lambda_4 \right ] \right \}  \ . \eeq

Our convention for the curvature scalar $R^{(D)}$ is such that, in a
weak field expansion, $R^{(D)} \sim g^{A\ \ B}_{\ \ A, \ B} -
g^{AB}_{\ \ \ ,AB}$, i.e. the sign of the curvature is opposite to the
sign of $R$. Then $\Lambda_4$ in (\ref{2.1e}) corresponds to a positive
cosmological constant for $\Lambda_4 > 0$. \par

Smoothing out the 3-brane corresponds to a replacement of $\delta^N(y)$
in (\ref{2.1e}) by a profile function $f(y)$. Let us introduce
spherical coordinates in the extra dimensions: $y^i = \{ r, y^{\alpha}
\}$ where $y^{\alpha}$, $\alpha = 1 \dots N-1$, are angles on
$S^{(N-1)}$ (an infinitely thin brane would be situated at $r = 0$).
The profile function now depends on $r$ only, $f = f(r)$. $f(r)$ is
non-vanishing only for $r \leq r_b$, where $r_b$ is the width of the
brane. \par

An explicit solution of Einstein's equation everywhere, i.e. also for
$r\leq r_b$, is not possible for general profile functions $f(r)$.
Therefore, as stated in the introduction, we make the particularly
simple choice $f(r) \sim \delta (r - r_b)$. The blown-up brane is then
``hollow'', all tension is located at the surface $r = r_b$. It
corresponds to an infinitely thin $D-2$ brane with a geometry
$R^{(3,1)} \times S^{(N-1)}$, where $S^{(N-1)}$ is a sphere with radius
$r_b$. The fact that this $D-2$ brane is again infinitely thin does not
necessarily induce singularities in its surrounding gravitational
field, since its co-dimension is just 1 (the dimension of its world
volume is $D-1$). \par

Correspondingly we should also replace the Einstein term
$\sqrt{-g^{(4)}} R^{(4)}$ on the brane by $\sqrt{-g^{(D-1)}}
R^{(D-1)}$, since this will be the structure induced by radiative
corrections \cite{3r}. The $D-1$ coordinates along the brane are
$x^{\mu}$ and the angles $y^{\alpha}$.  Thus the action (\ref{2.1e}) is
replaced by

\beq
\label{2.2e}
S = \int d^4x\ d^Ny \left \{ \sqrt{-g^{(D)}} {1 \over 2 \kappa_D}
R^{(D)} + \sqrt{- g^{(D-1)}} \delta (r - r_b) \left [ {1 \over 2
\kappa_{D-1}}  R^{(D-1)} + \Lambda_{D-1} \right ] \right \}  \ . \eeq

Subsequently we will be interested in Poincar\'e invariant (on
$R^{(3,1)}$) and spherically symmetric (on $S^{(N-1)}$) background
metrics with $g_{r\mu} = g_{r\alpha} = 0$; then we have
$\sqrt{-g^{(D)}} = \sqrt{-g^{(D-1)} g_{rr}}$. \par

The Einstein equations derived from (\ref{2.2e}) read

\beq
\label{2.3e}
{1 \over \kappa_D} G_{AB}^{(D)} = {1 \over \sqrt{g_{rr}}} \delta (r -
r_b) \left [ - {1 \over \kappa_{D-1}} G_{AB}^{(D-1)} + g_{AB}^{(D-1)}
\Lambda _{D-1} \right ]  \eeq

\noi where $g_{AB}^{(D-1)}$ is the pull-back of $g_{AB}^{(D)}$ onto the
brane, and the Einstein tensor $G_{AB}^{(D-1)}$ is constructed from
$g_{AB}^{(D-1)}$. For $A = r$ (or $B = r$) we have $G_{rB}^{(D-1)} =
g_{rB}^{(D-1)} = 0$. Up to now this is the same setup as considered in
\cite{10r}. \par

A static metric with the desired symmetries has the form

\beq
\label{2.4e}
ds^2 = A^2 (r) \ \eta_{\mu\nu} \ dx^{\mu} \ dx^{\nu} + B^2(r)\ dr^2 +
C^2(r) \ d\Omega_{N-1}^2 \ , \eeq

\noi where $d\Omega_{N-1}^2$ is the metric on $S^{(N-1)}$. Note that
our convention for $C(r)$ is such that flat space corresponds to $C(r)
= r$. The parameters $\kappa_{D-1}$, $\Lambda_{D-1}$ in (\ref{2.2e})
are then related to the original 4-dimensional parameters $\kappa_4$,
$\Lambda_4$, as

\bminiG{2.5e}
\label{2.5ae}
{1 \over \kappa_{D-1}} = {1 \over \Omega_{N-1} \ C^{N-1}(r_b)} \ {1
\over \kappa_4} \ , 
\eeeq
\beeq
  \label{2.5be}
\Lambda_{D-1} = {1 \over \Omega_{N-1} \ C^{N-1}(r_b)} \ \Lambda_4
\emini

\noi where $\Omega_{N-1}$ is the volume of the unit sphere $S^{(N-1)}$.
An asymptotically (for $r \to \infty$) flat solution of the empty space
Einstein equations (for $r > r_b$) for $A(r)$, $B(r)$ and $C(r)$ has
been given in \cite{13r,14r}:

\bminiG{2.6e}
\label{2.6ae}
A(r) = f_o^{-{1 \over 4} \sqrt{{N - 1 \over N+2}}}
\eeeq
\beeq
  \label{2.6be}
B(r) = f_o^{{1 \over N-2} \left ( \sqrt{{N - 1 \over N+2}}- {N - 3 
\over 2}\right
)} \eeeq
\beeq
  \label{2.6ce}
C(r) = r f_o^{{1 \over N-2} \left (  {1 \over 2} + \sqrt{{N - 1 \over 
N+2}}\right
)}
\emini

\noi with

\beq
\label{2.7e}
f_o = 1 + {\alpha \over r^{N-2}}\ .
\eeq

\noi (The index $o$ attached to $f_o$ stands for ``outside''.) The
parameter $\alpha$ in (\ref{2.7e}) can also be written as $\alpha = \pm
\ r_0^{N-2}$, and will be determined in terms of the parameters in the
action (\ref{2.2e}) below. \par

Next we seek for a solution of the empty space Einstein equations for 
$r < r_b$. In \cite{10r} the metric for $r < r_b$ has been chosen flat
(constant), and no singularity free solution to the junction conditions
(see below) has been found. \par

Instead, we chose for the metric for $r < r_b$ a copy of the metric 
for $r > r_b$ after a coordinate transformation

\beq \label{2.8e}
r' = {r_b^2 \over r}
\eeq

\noi and renaming $r' \to r$. The coefficient $r_b^2$ in (\ref{2.8e})
ensures that all components of the metric are continuous across $r =
r_b$. Explicitely the metric for $r < r_b$ reads

\bminiG{2.9e}
\label{2.9ae}
A(r) = f_i^{-{1 \over 4} \sqrt{{N - 1 \over N+2}}}
\eeeq
\beeq
  \label{2.9be}
B(r) = {r_b^2 \over r^2}\ f_i^{{1 \over N-2} \left ( \sqrt{{N - 1 
\over N+2}}- {N -
3 \over 2}\right )} \eeeq
\beeq
  \label{2.9ce}
C(r) = {r_b^2 \over r}\ f_i^{{1 \over N-2} \left (  {1 \over 2} + 
\sqrt{{N - 1
\over N+2}}\right )}
\emini

\noi with

\beq
\label{2.10e}
f_i = 1 + \alpha \left ( {r \over r_b^2}\right )^{N-2}
\eeq

\noi and hence $f_i(r_b) = f_o (r_b)$. \par

If one introduces an exponential radial coordinate $y$ through $r = r_b
\exp (y/r_b)$ for $r > r_b$, one sees that eqs. (\ref{2.9e}) and
(\ref{2.6e}) are related by a $\Z_2$-symmetry (reflection) around $y =
0$. For convenience (the junction conditions below) we stick to a
single coordinate $r$  ranging from 0 to $\infty$; the price to pay are
the coordinate singularities for $r \to 0$ in (\ref{2.9e}): The
divergence in $B(r)$ indicates that the point $r = 0$ is at infinite
proper distance from the ``surface'' at $r_b$ (since it is an image of
$+ \infty$), and the behaviour of $C(r)$ shows that the proper volume 
of spheres at radii $r < r_b$ blows up the same way as the one of
spheres at radii $r > r_b$. \par

In spite of the fact that the volumes both ``inside'' and ``outside''
are infinite (mirrors of each other), the proper volume of the extra 
dimensional part of the brane on $S^{(N-1)}$ at $r = r_b$ is finite and
given by $\Omega_{N-1} C^{N-1}(r_b)$. \par

Finally we have to show, however, that the metric (\ref{2.6e}) for  $r
> r_b$, and (\ref{2.9e}) for $r < r_b$, allows to satisfy the junction
conditions at $r = r_b$, i.e. to match the coefficients of $\delta (r -
r_b)$ in the Einstein equations (\ref{2.3e}). \par

Let us first discuss the coefficients of $\delta (r - r_b)$ on the
right-hand side of eq. (\ref{2.3e}), which are non-vanishing only for
$\{A,B\} = \{\mu , \nu\}$ or $\{ \alpha , \beta\}$. To this end we need
the $\{ \mu , \nu \}$, $\{\alpha , \beta \}$-components of the Einstein
tensor $G_{AB}^{(D-1)}$ computed from the pull-back metric
$g_{AB}^{(D-1)}$ \cite{10r} (Subsequently $A$, $B$ and $C$ denote
$A(r)$, $B(r)$, and $C(r)$ evaluated at $r = r_b$):

\bminiG{2.11e}
\label{2.11ae}
G_{\mu\nu}^{(D-1)}(r_b) = {A^2 (N-1) (N-2) \over 2 C^2} \ \eta_{\mu\nu}
\ ,
\eeeq
\beeq
  \label{2.11be}
G_{\alpha\beta}^{(D-1)} (r_b) = {1 \over 2} (N-2) (N-3)
g_{\alpha\beta}^{(N-1)}\ ,
\emini

\noi where $g_{\alpha\beta}^{(N-1)}$ is the metric on $S^{(N-1)}$ with
unit radius. Thus the non-vanishing right-hand sides of eq.
(\ref{2.3e}) read

  \bminiG{2.12e}
\label{2.12ae}
\delta (r - r_b) \eta_{\mu\nu} \left ( - {A^2(N-1)(N-2) \over 2
\kappa_{D-1} B C^2} + {A^2 \over B} \Lambda_{D-1} \right ) \ , \eeeq
\beeq
  \label{2.12be}
\delta (r - r_b) g_{\alpha\beta}^{(N-1)} \left ( - {(N-2)(N-3) \over 2
\kappa_{D-1} B } + {C^2 \over B} \Lambda_{D-1} \right ) \ .
\emini

\noi Terms $\sim \delta (r - r_b)$ appear on the left-hand side of eq.
(\ref{2.3e}) due to the discontinuous first derivatives of the metric 
across $r = r_b$. Generally one finds

\bminiG{2.13e}
\label{2.13ae}
G_{\mu\nu}^{(D)} = \delta (r - r_b) \eta_{\mu\nu} {A^2 \over B^2} \left
( - 3 \left [ {A'' \over A}\right ] - (N-1) \left [ {C'' \over C}\right
]  \right ) + \ \hbox{regular} \ , \eeeq  
\beeq
\label{2.13be}
G_{\alpha\beta}^{(D)} = \delta (r - r_b) g_{\alpha\beta}^{(N-1)} {C^2 
\over B^2} \left ( - 4 \left [ {A'' \over A}\right ] - (N-2) \left [
{C'' \over C}\right ] \right ) + \ \hbox{regular}
\emini

\noi where

\beq \label{2.14e}
\left [ {A'' \over A}\right ] = \left . \left ( {\partial_r A \over 
A}\right |_{r_b + \varepsilon} -  \left . {\partial_r A \over A} \right
|_{r_b - \varepsilon} \right )_{\varepsilon \to 0}\eeq

\noi and similarly for $[C''/C]$. No terms $\sim \delta (r-r_b)$ appear
in $G_{rA}^{(D)}$ for any $A$, in agreement with the right-hand side of
eq. (\ref{2.3e}). From $A(r)$, $C(r)$ in eqs. (\ref{2.6e}) and
(\ref{2.9e}) one finds

\bminiG{2.15e}
\label{2.15ae}
\left [ {A'' \over A} \right ] = {N-2 \over 2r_b}\left (1 - {1 \over
f}\right ) \sqrt{{N-1 \over N+2}} \ , \eeeq
\beeq
  \label{2.15be}
\left [ {C'' \over C} \right ] = {1 \over r_b}\left (1 - 2  \sqrt{{N-1
\over N+2}} + {1 \over f}{\left (1 + 2 \sqrt{{N-1 \over N+2}}\right
)}\right ) \ .
\emini

\noi where $f = f_o(r_b) = f_i (r_b)$. 
\noi Thus the terms $\sim \delta (r - r_b)$ in eq. (\ref{2.3e}) imply 

\bminiG{2.16e}
\label{2.16ae}
{A^2 \over \kappa_D B^2 r_b} \left (1-N+w + {1 \over f}(1-N-w)\right )
 = {A^2 \over B} \left ( - {(N-1)
(N-2) \over 2 \kappa_{D-1}C^2} + \Lambda_{D-1} \right ) \ ,\nn \\ 
\eeeq
\beeq
\label{2.16be}
-{C^2 \over \kappa_D B^2 r_b} (N-2)\left (1+{1 \over f} \right ) = {C^2
\over B} \left ( - {(N-2)(N-3) \over 2 \kappa_{D-1} C^2} +
\Lambda_{D-1} \right ) \ , \nn \\
\emini

\noi where we have defined $w = \sqrt{(N+2)(N-1)}/2$.  These equations
can be considerably simplified. Using $r_b B = C / \sqrt{f}$ (cf. eqs.
(\ref{2.6be}) and (\ref{2.6ce}), or (\ref{2.9be}) and (\ref{2.9ce}))
they can be brought into the form

\bminiG{2.17e}
\label{2.17ae}
\sqrt{f}(1-w) + {1 \over \sqrt{f}} (1+w) =  
{(N-2)\kappa_D \over \kappa_{D-1}C} \ , \eeeq
\beeq
\label{2.17be}
\sqrt{f}(N-1+(N-3)w)+{1 \over \sqrt{f}}(N-1-(N-3)w)
= -2\kappa_D \Lambda_{D-1} C \ . \nn \\  \emini

\noi Let us recall the basic parameters in the action (\ref{2.2e}) 
describing the
blown-up brane. These are \par

i) The $D = 4 + N$-dimensional gravitational constant $\kappa_D$, which
can be written in terms of a $D$-dimensional gravitational scale $M_D$
as

\beq
\label{2.18e}
{1 \over \kappa_D} = M_D^{2+N} \ .
\eeq

ii) The radial position of the $S^{(N-1)}$-part of the brane in the $N$
extra dimensions is given by $r_b$; however, only its proper volume 
$V_{S^{(N-1)}}$ is of physical significance, which is given by

\beq
\label{2.19e}
V_{S^{(N-1)}} = \Omega_{N-1} \ C^{N-1}
\eeq

\noi where $\Omega_{N-1}$ is the volume of the unit sphere.  Eq.
(\ref{2.19e}) explains the physical meaning of $C \equiv C(r_b)$.\par

iii) The gravitational constant $\kappa_{D-1}$ on the brane is related
through (\ref{2.5ae}) to the effective 4-dimensional gravitational
constant $\kappa_4$. Defining $\kappa_4 = M_{Pl}^{-2}$ we thus have

  \beq
\label{2.20e}
{1 \over \kappa_{D-1}} = {1 \over V_{S^{(N-1)}} \kappa_4} = {M_{Pl}^2
\over \Omega_{N-1} C^{N-1}} \ . \eeq

iv) The cosmological constant $\Lambda_{D-1}$ (or tension) on the brane
is related through (\ref{2.5ae}) to the 4-dimensional cosmological
constant $\Lambda_4$:

\beq
\label{2.21e}
\Lambda_{D-1} = {\Lambda_4 \over V_{S^{(N-1)}}} = {\Lambda_4 \over
\Omega_{N-1} C^{N-1}} \ .
\eeq

v) The width of the resulting gravitational field in the radial
direction is given by a parameter $\alpha$ in $f_i(r_b) = f_o(r_b)
\equiv f$, with

\beq
\label{2.22e}
f = 1 + {\alpha  \over r_b^{N-2}} \ .
\eeq

\noi Instead of $\alpha$, $f$ can be considered as a parameter to be
solved for. Then $r_b$ is given in terms of $C$ and $f$ through eq. 
(\ref{2.6ce}) at $r = r_b$, and $\alpha$ in terms of $f$ and $r_b$
through (\ref{2.22e}). \par

Now we have 2 equations (\ref{2.17e}) (originating from the $\{\mu ,
\nu\}$ and $\{ \alpha , \beta \}$ components of the junction
conditions) to solve. This implies immediately that the 4 input
parameters i) - iv) cannot be arbitrary; one ``fine-tuning'' relation
has to be  satisfied. Note that we count $C$ (or $r_b$) as an input
parameter: Its origin is the profile function in the radial direction
of the ``blown-up'' 3-brane (localized at its surface), introduced
originally as an UV regulator. \par

Furthermore some consistency conditions have to be satisfied: $C$ and
notably $f$ have to be positive. If $f < 0$, the metric is generically 
complex, and there exists always a naked singularity in the bulk
\cite{13r} which is precisely what we want to avoid. \par

Eqs. (\ref{2.17e}) can be solved most easily with $\kappa_D$, 
$\kappa_{D-1}$ and $C$ (positive) as input: Then eq. (\ref{2.17ae})
determines $f$, for which luckily always a positive solution exists.
Subsequently eq. (\ref{2.17be}) fixes $\Lambda_{D-1}$ (or $\Lambda_4$
with $C$ given). One sees immediately that $\Lambda_4$ has to
be fine-tuned. \par

Apart from this constraint on $\Lambda_4$ (or $\Lambda_{D-1}$), the 
existence of consistent (singularity-free) solutions is quite
remarkable, given the general presence of naked singularities in the
case of flat $p$-branes in $N$ codimensions \cite{13r}. \par

Let us first consider the case without an Einstein term on the brane,
i.e. $\kappa_{D-1}^{-1} \to 0$. Then eq. (\ref{2.17ae}) fixes $f =
(w+1)/(w-1)$, and eq. (\ref{2.17be}) gives $\kappa_D
\Lambda_{D-1}C = -2\sqrt{(N+2)(N-1)(N-2)/(N+3)}$. Thus, with eqs. 
(\ref{2.18e}) and (\ref{2.21e}),

\beq
\label{2.23e}
M_{\Lambda}^4 \sim M_D^{2+N} \ C^{N-2} \ .
\eeq

\noi With $f \sim {\cal O}(1)$ we have $r_0^{N-2}\ (\equiv \alpha)\ 
\sim r_b^{N-2} \sim C^{N-2}$, hence

\beq
\label{2.24e}
r_0^{N-2} \sim M_{\Lambda}^4 \ M_D^{-(N+2)} \ .
\eeq

\noi This order of magnitude of $r_0$ coincides with  the one in
\cite{13r,14r}.\par

Of phenomenological interest (cf. the next section) is, however, the
opposite limit

\beq
\label{2.25e}
{\kappa_D \over \kappa_{D-1} C} \gg 1 \ . \eeq

\noi Now eq. (\ref{2.17ae}) gives

\beq
\label{2.26e}
\left (\sqrt{f}\right )^{-1} \sim {\kappa_D \over \kappa_{D-1} C}\ .
\eeq

\noi ($f\ <\ 1$ implies that the parameter $\alpha$, introduced below
eq. (\ref{2.7e}), is negative.) Now eq. (\ref{2.26e}) implies, from eq.
(\ref{2.6ae}) or (\ref{2.9ae}) for $r = r_b$,

\beq
\label{2.27e}
A \sim \left ({\kappa_D \over \kappa_{D-1} C}\right )^{{1\over
2}\sqrt{{N-1\over N+2}}}
\eeq

As we will see in the next section, at the example of the scalar 
propagator, only in this limit the graviton propagator on the brane has
the  possibility to behave 4-dimensionally over a wide range of
scales.\par

\mysection{The Scalar Propagator}
\hspace*{\parindent}
As a first step towards the momentum dependence of the brane-to-brane
graviton propagator in the present gravitational background we will
study the corresponding propagator of a (dimensionless) scalar field,
whose kinetic terms in the bulk and on the brane have identical
coefficients as the  Einstein terms in the action (\ref{2.2e}).
Although an exact expression for this  propagator cannot be given, the
essential features of its momentum dependence can be  deduced using
results from \cite{14r}. \par

The differential equation for the general (bulk-to-bulk) scalar 
propagator has the form

\beq
\label{3.1e}
\left ( {1 \over \kappa_D} \sq^{(D)} + {1 \over \kappa_{D-1}B} \delta
(r - r_b) \sq^{(D-1)} \right ) G(x-x', y, y') = {- 1 \over
\sqrt{-g^{(D)}}}  \delta^4(x-x') \delta^N(y-y')\ . \eeq

We recall that $A(r)$, $B(r)$ and $C(r)$ define the metric
(\ref{2.4e}), and $A$, $B$ and $C$ denote these functions at $r = r_b$.
The $N$ extra coordinates $y^i$ are split into $y^i = \{r,
y^{\alpha}\}$ where $y^{\alpha}$ are  the angles on $S^{(N-1)}$, and
subsequently $g_{\alpha \beta}$ will denote the (dimensionless) angular
part of the metric on $S^{(N-1)}$. \par

Then the Laplacians in (\ref{3.1e}) are

\beq
\label{3.2e}
\sq^{(D)}  = A^{-2}(r) \partial_{\mu} \eta^{\mu\nu} \partial_{\nu} + {1
\over \sqrt{-g^{(D)}}} \partial_r \sqrt{-g^{(D)}} B^{-2}(r) \partial_r
+ C^{-2} (r) \partial_{\alpha} g^{\alpha\beta} \partial_{\beta} \ ,
\eeq

\noi and

\beq
\label{3.3e}
\sq^{(D-1)} = A^{-2} \partial_{\mu} \eta^{\mu\nu} \partial_{\nu} +
C^{-2} \partial_{\alpha} g^{\alpha\beta} \partial_{\beta} \ . \eeq

The determinant of the $D$-dimensional metric is

\beq
\label{3.4e}
\sqrt{-g^{(D)}} = A^4(r) \ B(r)\ C^{N-1}(r) \sqrt{\det (g_{\alpha
\beta})} \ .
\eeq

We will Fourier transform $G(x-x',y,y')$ with respect to the 4
Minkowski coordinates $x^{\mu} - x'^{\mu}$, whereupon $\partial_{\mu}
\eta^{\mu\nu}\partial_{\nu}$ is replaced by $-p^2$. Furthermore,
following \cite{14r}, we will decompose the angular part of the
propagator into spherical harmonics $Y_{\ell}^{m_a}$ ($a = 1 \dots
N-2$) on $S^{(N-1)}$:

\beq
\label{3.5e}
G(x-x',r,y^{\alpha}, r', y'^{\alpha}) = \int {d^4 p \over (2 \pi)^4}
e^{ipx} \Omega_{N-1} \sum_{\ell , m_a} Y_{\ell}^{m_a*}(y^{\alpha})
Y_{\ell}^{m_a}(y'^{\alpha})\cdot G_{\ell} (p,r,r')\eeq

\noi where $\Omega_{N-1}$ is the volume of the unit sphere, and the
spherical harmonics satisfy

\beq
\label{3.6e}
\sum_{\ell ,m_a} Y_{\ell}^{m_a*}(y^{\alpha}) \
Y_{\ell}^{m_a}(y'^{\alpha}) = {1 \over \sqrt{\det (g_{\alpha \beta})}}
\delta^{N-1} (y^{\alpha} -  y'^{\alpha} ) \ ,
\eeq

\beq
\label{3.7e}
\partial_{\alpha} g^{\alpha \beta} \partial_{\beta}\  
Y_{\ell}^{m_a}(y^{\alpha}) =
- \ell (\ell + N - 2) Y_{\ell}^{m_a} (y^{\alpha}) \ . \eeq

Putting $r' = r_b$, the equation for the bulk-to-brane propagator
$G_{\ell}(p,r,r_b)$ becomes

\bea
\label{3.8e}
&&\left \{ {1 \over \kappa_D} \left [ A^{-2}(r) p^2 - {1 \over
\sqrt{-g^{(D)}}} \partial_r \sqrt{-g^{(D)}} B^{-2} (r) \partial_r +
C^{-2} (r) \ell (\ell + N - 2) \right ] \right . \nn \\
&&\left . + {1 \over \kappa_{D-1} B} \delta (r- r_b) \left [ A^{-2} \
p^2 + C^{-2} \ell (\ell + N - 2)\right ] \right \} G_{\ell} (p, r, r_b)
\nn \\
&&= {1 \over A^4B C^{N-1} \Omega_{N-1}} \ \delta(r-r_b) \ .\eea

\noi Following \cite{5r} the solution of eq. (\ref{3.8e}) can be
written as

\beq
\label{3.9e}
G_{\ell}(p,r,r_b) = {D_{\ell}(p, r, r_b) \over 1 + {A^4 C^{N-1}
\Omega_{N-1} \over \kappa_{D-1}} \left ( A^{-2} p^2 + C^{-2} \ell (\ell
+ N-2)\right ) D_{\ell} (p, r_b, r_b)} \eeq

\noi where $D_{\ell} (p, r,r')$ is the propagator in the 
absence of field dependent terms on the brane and satisfies

\bea
\label{3.10e}
&&{1 \over \kappa_D} \left [ A^{-2} (r) p^2 - {1 \over \sqrt{-g^{(D)}}}
\partial_r \sqrt{-g^{(D)}} B^{-2}(r) \partial_r + C^{-2}(r) \ell (\ell
+ N-2)\right ] D_{\ell} (p,r,r') \nn \\
&&= {1 \over A^4 B C^{N-1} \Omega_{N-1}} \ \delta (r - r') \ .\eea

\noi Defining an operator ${\cal O}_{\ell}(r)$ as (using (\ref{3.4e})
for $\sqrt{-g^{(D)}}$)
\beq
\label{3.11e}
{\cal O}_{\ell}(r) = {-1 \over A^2(r) B(r)C^{N-1}(r)} \partial_r 
A^4(r) B^{-1}(r) C^{N-1}(r) \partial_r + {A^2(r) \over C^2(r)} \ell
(\ell + N-2) \eeq

\noi eq. (\ref{3.10e}) can be rewritten as

\beq
\label{3.12e}
\left ( {\cal O}_{\ell}(r) + p^2 \right ) D_{\ell}(p,r,r') = k \delta
(r-r')
\eeq

\noi with

\beq
\label{3.13e}
k = {\kappa_D \over A^2 B C^{N-1} \Omega_{N-1}} \ .
\eeq

The wave equation corresponding to eq. (\ref{3.12e}) (with a vanishing
right-hand side, and in the notation $p^2 = - m^2$) has been studied in
\cite{14r}. There the setup is slightly different than the one 
considered here: In \cite{14r} the brane (for the purpose of solving
the wave equation) is located at $r = 0$, and the functions $A(r)$,
$B(r)$ and $C(r)$ in ${\cal O}_{\ell}$ are of the form (\ref{2.6e}) for
$r > 0$. The remarkable result in \cite{14r} is that inspite
of the singular behaviour of these functions for $r \to 0$ the
solutions of the wave equation, and hence the scalar propagator, are
well defined. \par

In the present case the brane is located at $r = r_b$, and the
functions $A(r)$, $B(r)$ and $C(r)$ in ${\cal O}_{\ell}$ are of the
form (\ref{2.6e}) only for $r_b \leq r < \infty$, but of the form 
(\ref{2.9e}) for $0 < r \leq r_b$, hence $\Z_2$-symmetric under $r
\leftrightarrow r_b^2/r$. Asymptotically, both for $r \to \infty$ and
$r \to 0$, ${\cal O}_{\ell}(r)$ tends to the (radial) Laplacian in flat
space. Hence its spectrum is continuous, with eigenfunctions
$\phi(\ell,q,r)$ that behave as plane waves for $r \to \infty$ and $r
\to 0$. Here $q$ denotes a continuum variable conjugate to $r$. These
eigenfunctions can be decomposed into even ones under $\Z_2$ (with
vanishing radial derivatives at $r = r_b$) and odd ones under $\Z_2$
(which vanish at $r = r_b$).\par

It is possible to repeat the analysis in \cite{14r} of the operator 
${\cal O}_{\ell}(r)$ in the present setup (without singularities for $r
\to 0$, except for coordinate singularities, but jumps in the first
radial derivatives of $A(r)$, $B(r)$ and $C(r)$ at $r = r_b$), with the
same result: ${\cal O}_{\ell}(r)$ is self-adjoint and thus
semi-positive. Hence the eigenfunctions $\phi(\ell,q,r)$ satisfying

\beq
\label{3.14e}
{\cal O}_{\ell} \ \phi(\ell ,q, r) = e^2(q, \ell ) \ \phi (\ell , q, r)
\eeq

\noi with $e^2(q, \ell ) \geq 0$ represent a complete basis.\par

Although neither these functions nor their eigenvalues are not known
explicitely  (except for the zero modes \cite{14r}) they can be used to
construct a formal solution for the bulk Green function $D_{\ell}(p, r,
r')$ satisfying eq. (\ref{3.12e}):

\beq
\label{3.15e}
D_{\ell}(p, r, r') = k \int dq {\phi(\ell ,q, r) \ \phi(\ell, q, r')
\over p^2 + e^2 (q, \ell )} \eeq

The rough behaviour of $D_{\ell} (p, r, r')$ for $p \to 0$ can be
obtained from eq. (\ref{3.15e}) from the behaviour of the eigenvalues
$e^2(q, \ell )$ for $q \to 0$: First, dimensional analysis dictates
that we must have $e^2(q, \ell ) \sim q^2$. No powers of $r_b$ can
appear here, since the problem is non-singular for $r_b \to 0$
\cite{14r}. This shows already that $D_{\ell}(p \to 0, r, r') \sim
p^{-1}$. \par

Below we will be interested in the brane-to-brane propagator (and in
the $\ell = 0$ partial wave), and in this case we can be somewhat more
precise on the corresponding coefficient. Expanding ${\cal O}_{\ell =
0}$ around $r = r_b$ it becomes simply $- A^2 B^{-2}\partial_r^2$, in
which case (\ref{3.12e}) can be solved by simple Fourier transformation
and the argument of the $dq$-integral in (\ref{3.15e}) becomes $1/(p^2
+ A^2 B^{-2} q^2)$. Thus one finds

\beq \label{3.16e}
D_{\ell =0} (p \to 0, r_b , r_b) \sim {kB \over Ap}\ .
\eeq
 
Note that if one would compactify the extra dimensions inside a 
finite volume of size $\sim R^N$, the spectrum of ${\cal
O}_{\ell}(r)$ would be discrete. Then the integral $dq$ in
(\ref{3.15e}) would be replaced by a sum over discrete modes $q_i$
with spacing $\sim R^{-1}$, and the $p \to 0$ behaviour of $D_{\ell}$
would be entirely due to the zero mode: $D_{\ell} (p \to 0, r, r')
\sim k/(Rp^2)$ in contrast to the $p^{-1}$-behaviour obtained for
infinite extra dimensions. \par

The result (\ref{3.16e}) can be plugged into (\ref{3.9e}), together
with $k$ given in (\ref{3.13e}). Up to factors of ${\cal
O}(1)$ one finds

\beq
\label{3.17e}
G_{\ell} (p,r_b,r_b) \sim {\kappa_D\over A^3 C^{N-1}\Omega_{N-1}}
\cdot  {1 \over p + {A \kappa_D \over \kappa_{D-1}} \left ( A^{-2}
p^2 + C^{-2} \ell (\ell + N-2)\right )} \ . \eeq

Of particular importance is the denominator in the second factor in
(\ref{3.17e}), and the question where and when it behaves $\sim p^2$
such that a phenomenologically acceptable $p^2$ dependence of the
propagator is obtained. \par

First, to this end the first term has to be small with respect to the
second one, i.e.

\beq
\label{3.18e}
p \gg {A {\kappa_{D-1}} \over \kappa_D}
\eeq

\noi which is true at scales ${1 / p} \ll r_c$ with

\beq
\label{3.19e}
r_c \sim {\kappa_D \over A\kappa_{D-1}} \ .
\eeq

Now recall that the second and third terms in the denominator 
originate from the kinetic term $\sq^{(D-1)}$ localized on the brane,
whose extra  dimensions ($N-1$ of them) are compactified on
$S^{(N-1)}$. Once this term dominates, the propagator sees the
corresponding Kaluza-Klein states (partial waves  with $\ell > 0$) with
masses $M_{kk} \sim A/C$ in agreement with the metric  (\ref{2.4e}).
\par

Contributions of these states to the full propagator $G$ in
(\ref{3.5e}) (the sum over $G_{\ell}$) would not be acceptable, since
they would again modify its $p^2$-dependence and turn it into a $D = 4
+ N$ dimensional propagator. Thus one has to require

\beq
\label{3.20e}
p \ \lsim\ M_{kk} \sim {A \over C}
\eeq

\noi in which regime only the $\ell = 0$ mode contributes. \par

Once the inequalities (\ref{3.18e}) and (\ref{3.20e}) are satisfied, 
the $\ell = 0$ brane-to-brane propagator behaves as

\beq
\label{3.21e}
G_0 (p, r_b, r_b) \sim{\kappa_{D-1} \over C^{N-1} \Omega_{N-1}A^2p^2} 
\sim {\kappa_4 \over A^2 p^2} \ ,
\eeq

\noi where we have used (\ref{2.5ae}) in the last step. The  coefficient
coincides with the one obtained from a 4-dimensional action

\beq
\label{3.22e}
S = {1 \over 2\kappa_4} \int d^4x \sqrt{-g^{(4)}} \partial_{\mu} \phi 
g^{\mu\nu} \partial_{\nu} \phi \eeq

\noi with $g_{\mu\nu}^{(4)} = A^2 \eta_{\mu\nu}$, as it should. \par

The two inequalities (\ref{3.18e}) and (\ref{3.20e}) are compatible
only if

\beq
\label{3.23e}
{\kappa_D \over \kappa_{D-1}C} \gg 1\ .
\eeq

Let us consider the required orders of magnitude in more detail. For
the scale $r_c$, beyond which gravity would change, one can only
tolerate $r_c \ \gsim\ H_0^{-1}$ \cite{2r}, where $H_0$ is the Hubble
constant today. Hence, rewriting eq. (\ref{3.19e}), 

\beq
\label{3.24e}
r_c \sim {\kappa_D \over C\kappa_{D-1}} {C \over A} \ \gsim \ 
H_0^{-1} \ .
\eeq

\noi On the other side, in order not to change gravity at distances
down to $10^{-4}$m through Kaluza-Klein states with masses $M_{kk}$
trapped on $S^{(N-1)}$ we need

\beq
\label{3.25e}
M_{kk} \sim {A \over C} > 10^{-3}\ {\rm eV} \ .
\eeq

\noi With $H_0 \sim 10^{-33}$~eV eqs. (\ref{3.24e}) and (\ref{3.25e})
imply

\beq
\label{3.26e}
{\kappa_D \over C\kappa_{D-1}} > 10^{30} \ .
\eeq

\noi Now recall that, from the junction conditions derived in the
previous chapter, the combination of parameters in (\ref{3.26e}) is
related to the warp factor $A$ on the brane. From eqs. (\ref{2.26e}),
(\ref{2.27e}) with (\ref{3.26e}) one finds

\beq
\label{3.27e}
A > 10^{15\sqrt{{N-1\over N+2}}}\ .
\eeq

A huge warp factor on the brane aggravates the hierarchy problem (i.e.
the smallness of the Weak Scale as compared to the Planck Scale) instead
of solving it in the sense of \cite{15r} via a tiny warp factor on the
brane.\par

If one is willing to swallow this complication, one can translate the
constraint (\ref{3.26e}), via eqs. (\ref{2.18e}) and (\ref{2.20e}), into
a constraint on the $D$-dimensional gravitational scale $M_D$ and $C$,
which is related to the size of the blown-up brane in the extra
dimensions via (\ref{2.19e}). Then (\ref{3.26e}) gives 

\beq
\label{3.28e}
{M_{Pl}^2\over \Omega_{N-1} M_D^2 (M_D C)^N} > 10^{30}\ .
\eeq

\noi For $C \sim M_D^{-1}$ one obtains

\beq
\label{3.29e}
M_D < 10^{-15} M_{Pl}
\eeq

\noi as in the co-dimension 1 case in \cite{2r}. However, eq.
(\ref{3.28e}) allows to generate a huge number more easily, if one
allows for $C \ll M_D^{-1}$. In the extreme case $C \sim M_{Pl}^{-1}$
one obtains

\beq
\label{3.30e}
\left ( {M_{Pl} \over M_D} \right ) ^{N+2} > 10^{30}
\eeq

\noi which, for $N$ large, allows for much more modest ratios
$M_D/M_{Pl}$.\par

What would be the required tension (cosmological constant) on the brane?
From eqs. (\ref{2.17e}) one finds that, in the limit (\ref{2.25e}) and
(\ref{2.26e}), its right hand sides are of comparable magnitude. Using
eqs. (\ref{2.20e}) and (\ref{2.21e}) this gives

\beq
\label{3.31e}
\Lambda_4 \sim {M_{Pl}^2 \over C^2}
\eeq

\noi or $\Lambda_4 \sim M_{Pl}^2\ M_D^2$ in the case (\ref{3.29e}),
$\Lambda_4 \sim M_{Pl}^4$ in the case (\ref{3.30e}).\par

In order to see whether this scenario is phenomenologically viable, the
complete spectrum of physical fluctuations would have to be
studied.\par

The tensor structure of 4-dimensional gravity, in the presence of an
additional Einstein term on the brane, has recently been the subject of
numerous investigations [3, 9, 16-21]. Note that the problematic result
(ghosts and tachyons for $N \geq 2$) obtained in \cite{21r} cannot be
applied here, since we have a non-trivial gravitational background and,
notably, replaced $R^{(4)}$ by $R^{(D-1)}$ on the blown-up brane, hence
we have an effective codimension one scenario. The recent results in
the co-dimension 1 case, on the other hand, point towards a
phenomenologically acceptable tensor structure of 4-dimensional gravity
[16-20]. Nevertheless the details in the present case, as well as the
spectrum of 4-dimesional scalars (moduli), remain to be studied.\par

Independently thereof we want to emphasize that the present brane
configuration can also be studied in {\it compact} extra dimensions, in
which case the correct structure of 4-dimensional gravity is guaranteed
even without an additional Einstein term on the brane. Given the
present negative result on the solution of the cosmological constant
problem, and the required hierarchies of scales in the case of
infinite-volume extra dimensions (notably the need for $M_D \ll
M_{Pl}$, although our eq. (\ref{3.30e}) reduces the required
hierarchy), compact extra dimensions are perhaps somewhat more
promising. Work in this direction is in progress. 
\par \vskip 2 truecm

\noi {\large\bf Acknowledgement:} We thank Ch. Charmousis for numerous
stimulating discussions.

\end{document}